\documentclass[12pt]{iopart}

\usepackage{graphicx}

\begin{document}

\title[Noise-based signal processing]{Signal processing in local neuronal circuits based on activity-dependent noise and competition}

\author{Vladislav Volman$^{1,2}$ and Herbert Levine$^{1}$}
\address{
1. Center for Theoretical Biological Physics, University of California at San Diego, La Jolla, CA 92093-0319, USA \\
2. Computational Neurobiology Laboratory, The Salk Institute for Biological Studies, La Jolla, CA 92037, USA }

\begin{abstract}
We study the characteristics of weak signal detection by a recurrent neuronal network with plastic synaptic coupling. It is shown that in the presence of an asynchronous component in synaptic transmission, the network acquires selectivity with respect to the frequency of weak periodic stimuli. For non-periodic frequency-modulated stimuli, the response is quantified by the mutual information between input (signal) and output (network's activity), and is optimized by synaptic depression. Introducing correlations in signal structure resulted in the decrease of input-output mutual information. Our results suggest that in neural systems with plastic connectivity, information is not merely carried passively by the signal; rather, the information content of the signal itself might determine the mode of its processing by a local neuronal circuit.
\end{abstract}

\bibliographystyle{unsrt}

\maketitle
\baselineskip 18pt

\textbf{Information in neuronal circuits is represented by series of action potentials that are generated by neurons following the summation of many synaptic stimuli. Importantly, the strength of synaptic connection between a pair of neurons can be plastically adjusted in a manner that depends on the context of network's activity. In addition to the fast response to the action potential, synaptic transmission often has another component (asynchronous release), that can persist for $>100~msec$ following strong stimulation of the synapse. This constitutes a challenging example of activity-dependent noise. Given the fact that neuronal population is often organized in strongly connected local recurrent networks, and the wide spectrum of neuronal firing frequencies (from $1~Hz$ to $200~Hz$) it is of interest to ask how synaptic plasticity, and in particular activity-dependent asynchronous release, affect processing of different signals by these local circuits. Understanding the role of synaptic plasticity in the detection of weak stimuli can help to unveil the principles of information processing in neural systems.}

\section{Introduction}
Neurons exchange information with their peers by sending action potentials that are transmitted by chemical synapses with activity-dependent strength and are summed at the post-synaptic cell body to determine the temporal pattern of spike firing \cite{KochBook}. The observed rate of neuronal activities spans a wide spectrum, ranging from values as low as $1~Hz$ for spontaneous activity, to values as high as $\approx~200~Hz$ for fast rhythms observed in hippocampal regions \cite{Bragin95}. Considering that in many cases (for example in the CA3 region of hippocampus) neurons are organized in strongly connected local recurrent networks ($\approx~100$ neurons) that receive information from more distant parts of the brain \cite{Andersen71}, it is natural to inquire how these local circuits of nonlinear neurons with activity-dependent coupling process signals of different rhythmicity.\\
\indent~The effect of stimulus characteristics on the ability of a coupled neuronal ensemble to detect weak signals had been investigated earlier by several researchers using the conceptual framework of stochastic resonance (SR) \cite{Collins96,Bulsara96,Liu99}. Those studies assumed the coupling between the neurons to be static, or at best proportional to the voltage gradient. However, neurons usually exchange information via chemical synapses that can plastically adjust their strength in an activity-dependent manner on a variety of time-scales (from milliseconds to minutes). On the single synapse level, depression acts as a low-pass filter, letting through only signals with a typical time scale slower than the recovery from depression \cite{Fortune01}. In addition, synaptic transmission can often be characterized by two components: fast and strong phasic release of transmitter in response to the action potential that invades the synapses, and slow weaker asynchronous release (AR) that can persist for $>100~msec$ following strong stimulation of the synapse \cite{Goda94,Hagler01,Lau05}. Since asynchronous release is constrained by the total amount of available synaptic resource, this kind of "synaptic noise" is expected to have temporal structure that reflects prior activity, therefore constituting an interesting example of activity-dependent noise. It is also worth mentioning that memory effects on weak signal detection in the stochastic resonance regime were studied earlier for a bistable system with internal colored noise \cite{Neiman96}, and it was found that memory (an increase in noise correlation time) usually acted to suppress stochastic resonance.\\
\indent~In the present study, we aim to achieve a somewhat broader goal by asking the following question - how would the SR-like response of a \textit{plastic system with activity-dependent noise} depend on the characteristics of an input signal. Different facets of this problem have been addressed separately before (see, e.g. \cite{Liu99,Balenzuela05}), but none of these earlier studies investigated the consequences of an interaction between signal properties, presynaptic plasticity and in particular the effect of asynchronous component of synaptic transmission. We show here that when synaptic dynamics are endowed with plasticity and competing modes of signal transmission (evoked vs. asynchronous), the recurrent network acquires selectivity with regard to the properties of weak stimuli. Such stimulus selectivity can be modulated by changing the level of asynchronous release at the model network's synapses. Our results suggest that the dynamical control of AR in a local circuit (exerted, for example, by glial cells) could act to switch the attention of that local circuit to certain stimuli.

\section{Methods}
\begin{figure}
\centerline{
\resizebox{0.7\textwidth}{!}{\includegraphics{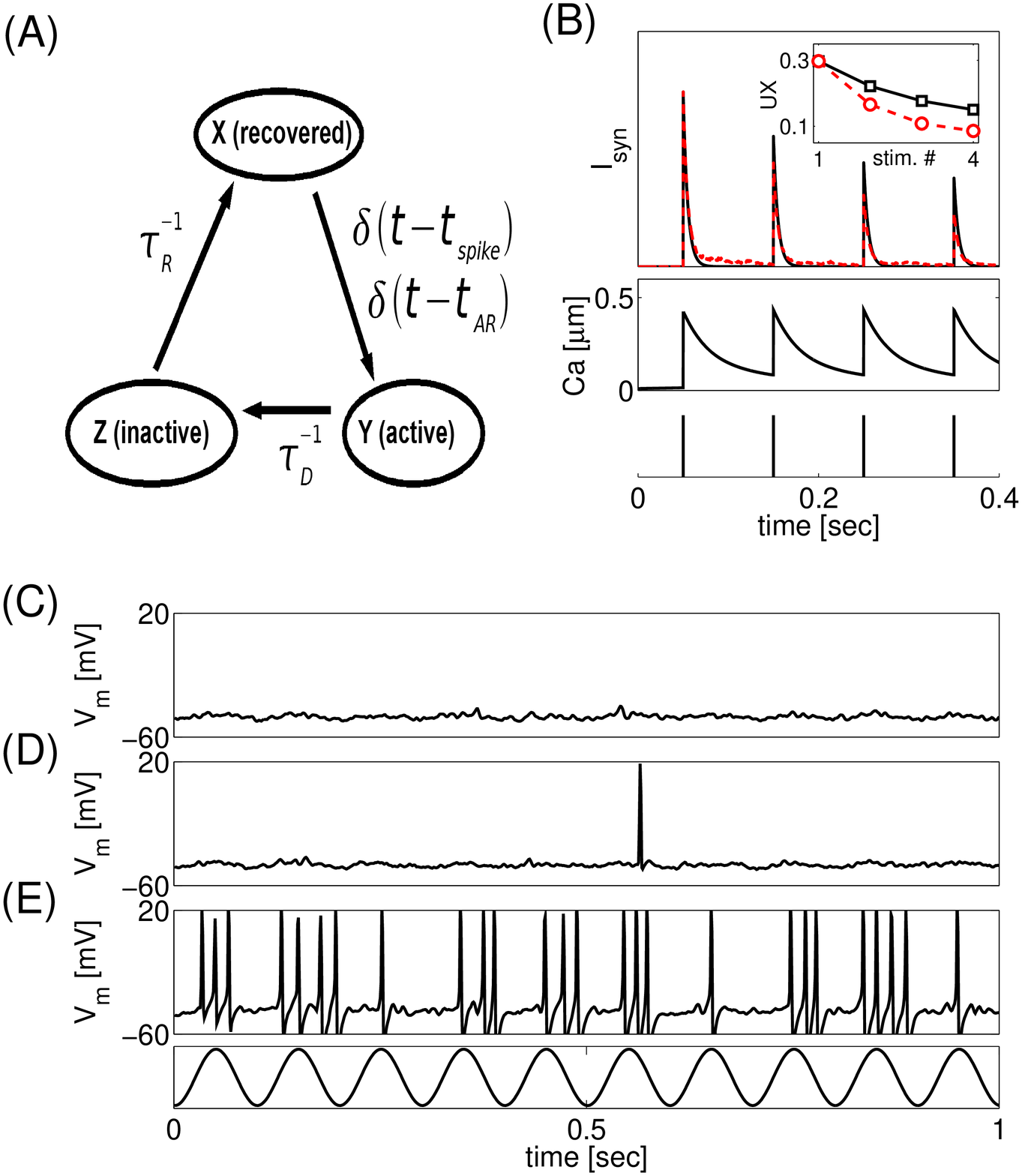}}}
\caption{\footnotesize Asynchronous release of neuro-transmitter from model synaptic terminals facilitates response to weak sub-threshold stimuli. A) Schematic presentation of the model for synaptic kinetics used here. B) Response of the model synapse to a repetitive series of spikes. Top: post-synaptic response (arbitrary units) for a model without AR (solid line), and with AR (dashed line). Asynchronous release acts to reduce synaptic response to spikes (inset). Middle: Pre-synaptic calcium trace. Bottom: series of spikes that were used to feed into the synapse. C,D,E) Examples of membrane potentials for a neuron that is subject to weak sub-threshold periodic stimulation, for the case of an isolated neuron (C), coupled to a network with the synchronous release only (D), and coupled with both synchronous and asynchronous release components (E).}
\label{fig1:Figure1Label}
\end{figure}

The neuronal and synaptic models employed here are the same one that were used in our previous studies \cite{Volman08}, and are based on slight modifications of the model that had been developed to study the characteristics of evoked reverberatory responses in cultured hippocampal networks \cite{Volman07,Lau05}.\\
\textit{Network model: }Throughout this study, unless otherwise indicated, the system size was taken to be $N=100$. For each pair $(i,j)$ of model neurons, an uni-directional connection was defined with the probability $p=0.1$. Self-connections were excluded. In this setup, the probability for a neuron to have $k$ incoming contacts is given by binomial formula $P(k,N,p)=\frac{(N-1)!}{k!(N-1-k)!}p^{k}(1-p)^{N-1-k}$. The values for network size and connectivity were chosen on the basis of existing evidence from hippocampal cultures \cite{Lau05}, to comply with the notion of "local circuit".\\
\textit{Neuronal model: }Information that arrives from synaptic and ionic channels is integrated by a neuron to yield a spike-time series. Here, we describe the neuron as a conductance-based one-compartment entity, using a modified version of the Morris-Lecar model \cite{MorrisLecar81,Prescott06}. This level of modeling represents a compromise between detailed multi-compartmental models that encompass realistic dendritic morphologies on one hand, and, on the other hand, over-simplified models of the integrate-and-fire variety. The ionic current through the neuronal membrane is modeled as -
\begin{equation}\label{eq:Eq_Neuro1}
I_{ion} = g_{Na}m_{\infty}(V-E_{Na})+g_{K}w(V)(V-E_{K})+g_{leak}(V-E_{leak})
\end{equation}
\begin{equation}\label{eq:Eq_Neuro2}
\dot{w} = 0.15(w_{\infty}(V)-w(V))cosh((V-V_{3})/2V_{4})
\end{equation}
\begin{equation}\label{eq:Eq_Neuro3}
m_{\infty} = 0.5(1+tanh((V-V_{1})/V_{2}))
\end{equation}
\begin{equation}\label{eq:Eq_Neuro4}
w_{\infty} = 0.5(1+tanh((V-V_{3})/V_{4}))
\end{equation}
With equations \ref{eq:Eq_Neuro1}-\ref{eq:Eq_Neuro4}, the dynamics of neuronal membrane potential are described as -
\begin{equation}
\label{eq:Eq_NeroMembrane}
C\dot{V} = -I_{ion}(t)+I_{bg}(t)-(V(t)-E_{R})\Sigma\bar{g}_{j}Y_{j}(t)+I_{signal}(t)
\end{equation}
with the term $\Sigma\bar{g}_{j}Y_{j}(t)$ representing summation over all incoming synaptic connections, $\bar{g}\in[0.5,0.8]~mS/cm^{2}$, and $Y_{j}(t)$ being the (time-dependent) strength of synapse from $j$-th neuron, modeled as described below. We consider a network with excitatory coupling, and therefore set synaptic reversal potential to $E_{R}=0~mV$. The term $I_{bg}$ is a background current that represents the summation of a large number of synaptic stimuli from neurons that are not part of the specific local circuit. This current is described by the Langevin equation $\dot{I}_{bg}=-I_{bg}/\tau_{n}+\sqrt{D/\tau_{n}}\mathcal{N}(0,1)$, with
$\tau_{n}=10~msec$, $D=0.64\cdot10^{-2} \mu A^{2}/cm^{4}$, and $\mathcal{N}(0,1)$ being uncorrelated Gaussian noise with zero mean and unitary variance. The term $I_{signal}(t)$ is the weak periodic external signal, of amplitude $max(I_{signal}(t))=1\frac{nA}{cm^{2}}$, with its frequency $\nu_{in}$ selected as explained.\\
\indent~The following parameter values were used: $E_{Na}=50~mV, E_{K}=-100~mV, E_{leak}=-55.8~mV, V_{1}=-1.2~mV, V_{2}=23~mV, V_{3}=-2~mV, V_{4}=21~mV, g_{Na}=10~mS/cm^{2}, g_{K}=10~mS/cm^{2}, C=1~\mu~F/cm^{2}$. Leak conductance of neuronal membrane was set to $g_{leak}=1.5~mS/cm^{2}$. With this choice of parameters, transition from quiescence to a spiking state is accomplished through a Hopf bifurcation.\\
\textit{Synaptic model: }Rather than attempting a complete biophysical description of the complex synaptic machinery that would include the quantal nature of vesicular release and sensitivity to spatial Calcium profiles, we use a phenomenological model that describes the synchronous activation of several active zones \cite{Volman07}. In this "mean-field" description, the kinetics of synaptic neurotransmitter resource are given by following equations -
\begin{equation}\label{eq:Eq_SynModel1}
\dot{X}_{j} = \frac{Z_{j}}{\tau_{R}}-X_{j}(U\delta(t-t_{s}^{j})+\xi\delta(t-t_{a}^{j}))
\end{equation}
\begin{equation}\label{eq:Eq_SynModel2}
\dot{Y}_{j} = \frac{-Y_{j}}{\tau_{D}}+X_{j}(U\delta(t-t_{s}^{j})+\xi\delta(t-t_{a}^{j}))
\end{equation}
\begin{equation}\label{eq:Eq_SynModel3}
\dot{Z}_{j} = \frac{Y_{j}}{\tau_{D}}-\frac{Z_{j}}{\tau_{R}}
\end{equation}
\begin{equation}\label{eq:Eq_SynModel4}
\eta(c) = \eta_{max}\frac{c^4}{c^4+K_{a}^{4}}
\end{equation}
\begin{equation}\label{eq:Eq_SynModel5}
\dot{c} = \frac{-\beta c^{2}}{c^{2}+K_{c}^{2}}+\gamma
log(\frac{c_{o}}{c})\delta(t-t_{s}^{j})+I_{p}
\end{equation}
Equations \ref{eq:Eq_SynModel1}-\ref{eq:Eq_SynModel5} describe the response of a $j$-th model synapse to action potentials that occur at times $t_{s}^{j}$. It is assumed that synaptic resource is either in recovered ($X$), active ($Y$) or inactive ($Z$) states, and therefore, as is seen from Equations \ref{eq:Eq_SynModel1}-\ref{eq:Eq_SynModel5}: $X+Y+Z=1$. The $Y\rightarrow Z$ and $Z\rightarrow X$ transitions are with rates $\tau_{D}^{-1}$ and $\tau_{R}^{-1}$ correspondingly, satisfying $\tau_{D}^{-1}\gg\tau_{R}^{-1}$. In addition to fast phasic transmission (the strength of which is modeled here as $UX_{j}$), there are asynchronous synaptic events of amplitude $\xi X_{j}$ that are generated at times $t_{a}^{j}$. AR event generation (resulting in the $t_{a}^{j}$) is treated as a Poisson process with a time-dependent rate $\eta(c)$, which depends on the synaptic Calcium concentration, $c$. The Calcium concentration increases due to action potentials, in proportion to the electro-chemical gradient across the synaptic membrane, and decays nonlinearly due to active pumping. Note that Eqns.\ref{eq:Eq_SynModel2},\ref{eq:Eq_SynModel4} are coupled with Eq.\ref{eq:Eq_SynModel5} through the spike times $t_{s}^{j}$ and $t_{a}^{j}$. The term $I_{p}$ ensures that in the absence of any presynaptic spikes, Calcium is maintained at a non-zero steady-state level. The essentials of the synaptic model are summarized in Figure \ref{fig1:Figure1Label}.\\
The following parameter values were used to model the properties of synaptic transmission: $\tau_{D}=5~msec, \tau_{R}=0.6~sec, K_{a}=0.1~\mu M, K_{c}=0.4~\mu M, \beta=2~\mu M/sec, \gamma=80~nM, c_{0}=2~mM, I_{p}=0.11~\mu M/sec, \xi=10^{-3}$. In all simulations, we set $U=0.4$.\\
\textit{Methods of analysis: }The output of a neuron is characterized by a sequence $\{t_{i}\}$ of spikes. An equivalent representation is by means of inter-spike-interval (ISI) series, defined as $ISI_{n}=t_{n+1}-t_{n}, n\geq1$. The dynamics of neuronal network is usually visualized by means of raster plot, in which each row marks the times of individual neuron firings ($y$ axis runs over all the neurons in the network). For signals of fixed frequency $\nu_{in}$, the response of a network is conveniently captured by the coherence of spiking (COS) measure \cite{COSRef}. The COS measure, $C_{S}$, is defined as $C_{S}\equiv\frac{N(0.9T~\leq ISI\leq~1.1T)}{N(ISI)}$, that is, the fraction of spikes that are within $20\%$ from the stimulation period, $T=\nu_{in}^{-1}$. To measure the variability of inter-spike interval (ISI) series, we compute its coefficient of variation, $CV(ISI)$, defined as $CV(ISI)\equiv\frac{\sigma(ISI)}{\mu(ISI)}$, where $\mu(ISI)$ is averaged ISI (inverse of firing rate), and $\sigma(ISI)$ is its standard deviation. All results shown are averages over 50 independent realizations.\\

\section{Results}
\subsection{Synaptic depression and asynchronous release optimize the detection of weak periodic stimuli}
\begin{figure}
\centerline{
\resizebox{0.7\textwidth}{!}{\includegraphics{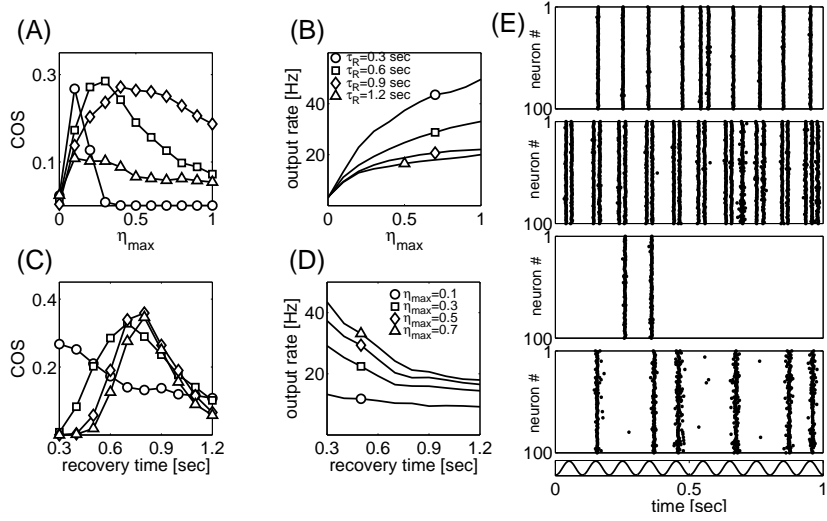}}}
\caption{\footnotesize Efficient detection of weak stimuli depends on the matching between asynchronous release and synaptic depression. A) Coherence of spiking plotted vs. rate of asynchronous release, for different values of depression recovery time: $\tau_{R}=0.3~sec$ (circles), $\tau_{R}=0.6~sec$ (squares), $\tau_{R}=0.9~sec$ (diamonds), $\tau_{R}=1.2~sec$ (triangles). B) Output rates plotted vs. AR rate, for different values of $\tau_{R}$ shown in (A). C) Coherence of spiking vs. $\tau_{R}$, for different values of AR rate: $\eta_{max}=0.1$ (circles), $\eta_{max}=0.3$ (squares), $\eta_{max}=0.5$ (diamonds), and $\eta_{max}=0.7$ (triangles). D) Output rates for all cases shown in (C). E) Examples of network's collective activity for different values of $\eta_{max}$ and $\tau_{R}$. From top to bottom: $\eta_{max}=0.1, \tau_{R}=0.3~sec$; $\eta_{max}=0.4, \tau_{R}=0.3~sec$; $\eta_{max}=0.1, \tau_{R}=1.2~sec$; $\eta_{max}=0.4, \tau_{R}=1.2~sec$. In all cases (A-D), the signal frequency is $\nu_{in}=10~Hz$.}
\label{fig2:Figure2Label}
\end{figure}

The goal of our study is to understand how short-term presynaptic plasticity (as manifested by synaptic depression) and activity-dependent asynchronous release (AR) of neurotransmitter affect the detection and processing of weak signals by a local neuronal circuit. To this end, we first investigated the dependence of network activity on the recovery time from presynaptic short-term depression and on the different AR levels, for a given pattern of signal. Detection here is monitored by the coherence of output spikes with the incoming periodic signal as captured by the COS measure. Results of this analysis, shown in Figure \ref{fig2:Figure2Label}, suggest that the efficient detection of weak periodic stimuli requires matching between synaptic depression and asynchronous release. Fast recovery time and high levels of AR increase the firing rate (Figure \ref{fig2:Figure2Label}B, circles) which leads to almost zero coherence for high AR rates due to the increased number of spikes fired during each signal cycle (Fig.\ref{fig2:Figure2Label}A, circles, and Fig.\ref{fig2:Figure2Label}E, second panel). On the other hand, a very strong depression reduces the firing rate (Fig.\ref{fig2:Figure2Label}B, triangles) and also reduces the ability to fire coherently with the signal (Fig.\ref{fig2:Figure2Label}A, triangles, and Fig.\ref{fig2:Figure2Label}E). Note also that the case of $\tau_{R}=1.2~sec$ is qualitatively different because in this regime of strong depression both the detection and the amplification of weak signal are compromised due to the low availability of synaptic resource. In between the two extremes of fast and slow recovery from depression, and for intermediate levels of AR, there exists a regime that yields high levels of spiking coherence. This is best seen from Fig.\ref{fig2:Figure2Label}C, where we plot the COS measure as a function of $\tau_{R}$, for different levels of AR. Note that low levels of AR (Figure \ref{fig2:Figure2Label}C, circles) result in a monotonic decrease of coherence as the value of $\tau_{R}$ is increased, i.e. moderate-to-strong expression of asynchronous release is needed to create an optimal recovery time. 

\subsection{Selectivity with respect to periodic stimuli}
\begin{figure}
\centerline{
\resizebox{0.7\textwidth}{!}{\includegraphics{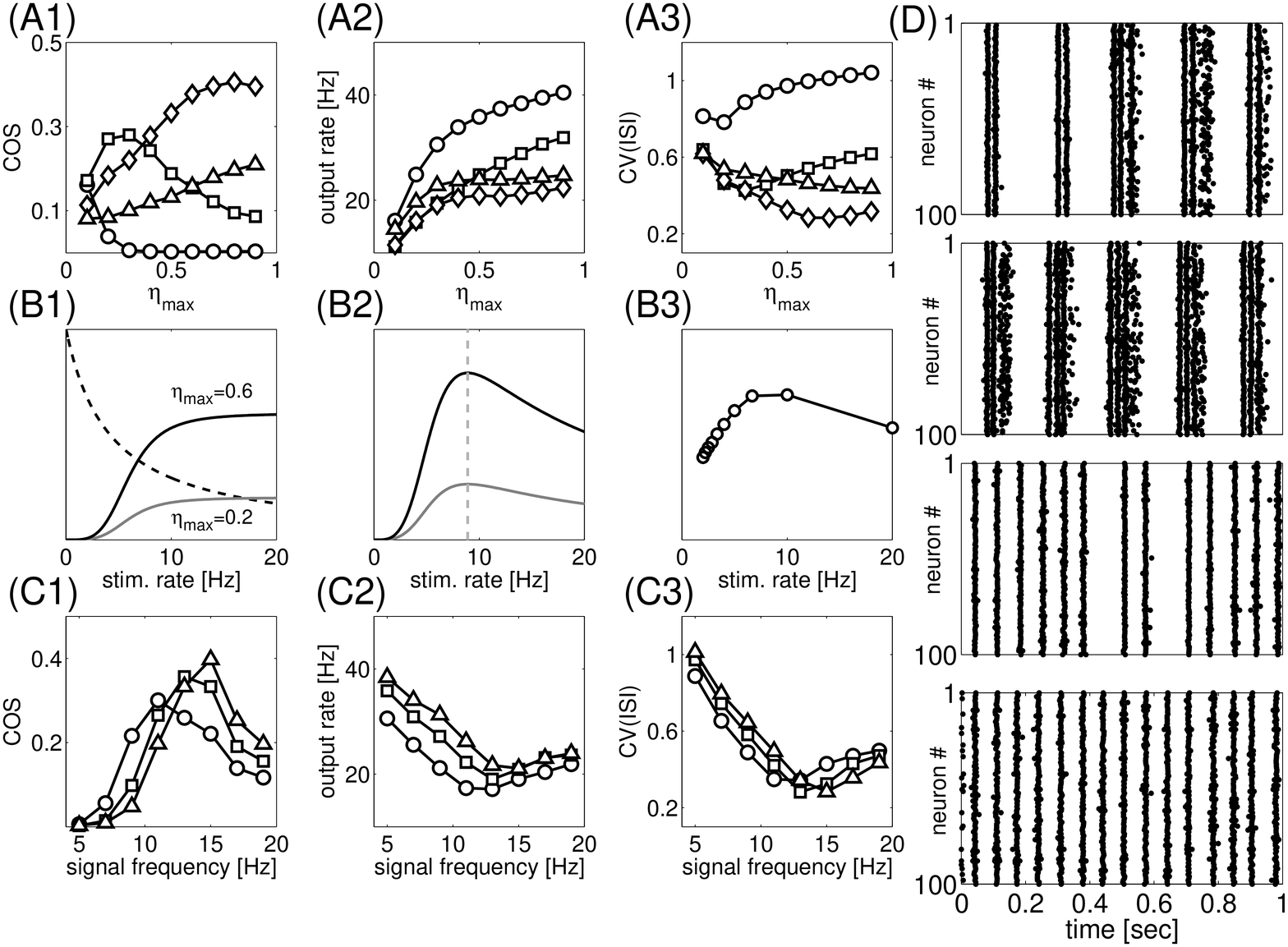}}}
\caption{\footnotesize Network's detection of weak periodic stimulation depends on the stimulus frequency. A1) Coherence of spiking vs. rate of asynchronous transmission, for periodic stimuli of different rates: $\nu_{in}=5~Hz$ (circles), $\nu_{in}=10~Hz$ (squares), $\nu_{in}=15~Hz$ (diamonds), $\nu_{in}=20~Hz$ (triangles). The amplitude of stimuli is the same for all cases. A2,A3) Averaged output firing rates (A2), and coefficients of inter-spike-interval series variation (A3), plotted vs. rate of asynchronous release, for all 4 cases shown in (A1). B1) Schematic presentation for steady state levels of available synaptic resource, $X$ (dashed line), and rates of asynchronous release (solid lines), plotted vs. the constant stimulation rate, $R$. B2) Synaptic drive due to asynchronous release, as in (B1), peaks for an optimal stimulation rate. B3) Asynchronous component of synaptic current vs. the stimulation rate of model synapse. C1) Coherence of spiking, plotted vs. stimulus frequency, for different levels of asynchronous transmission: $\eta_{max}=0.3$ (circles), $\eta_{max}=0.5$ (squares), $\eta_{max}=0.7$ (triangles). C2,C3) Averaged output firing rates (C2), and coefficients of ISI series variation (C3), plotted vs. $\eta_{max}$, for all four cases shown in (C1). D) Examples of network's collective activity for different levels of AR, and subject to periodic stimuli of different frequencies: $\eta_{max}=0.3, \nu_{in}=5~Hz$ (D1), $\eta_{max}=0.6, \nu_{in}=5~Hz$ (D2), $\eta_{max}=0.3, \nu_{in}=15~Hz$ (D3), $\eta_{max}=0.6, \nu_{in}=15~Hz$ (D4). For all cases (A-D), $\tau_{R}=0.6~sec$.}
\label{fig3:Figure3Label}
\end{figure}

We proceed to investigate the response of a network to properties of the weak periodic stimuli. Figure \ref{fig3:Figure3Label} summarizes the dependence of network's response on different periodic stimuli. As is seen, for a given set of parameters, the coherence of response to the $\nu_{in}=5~Hz$ stimulus is the lowest one, and actually decreases as the level of AR is increased. For intermediate input rate ($\nu_{in}=10~Hz$) the COS measure peaks at an optimal value of AR, attaining low values for low and high levels of $\eta_{max}$. For still higher input frequencies, coherence of output activity increases monotonically for all values of AR under consideration; however, the maximal values of COS measure become smaller for higher input rates (diamonds vs. triangles in Figure \ref{fig3:Figure3Label}A1). This relative decrease of coherence for high stimulation rates is a direct consequence of synaptic depression that causes the neuron to skip some cycles. On the other hand, for periodic signals that have low input frequency (for example, $\nu_{in}=5~Hz$), the "dwell time" (per cycle) near the spike generation threshold is increased (as compared with the same-amplitude signal of higher frequency). As a result, there can occur an increased number of spikes per cycle, and the stronger depression that follows after such an intense period of activity acts to reduce the coherence of output activity. This effect for a low-frequency stimulus correlates with a high output firing rate (Figure \ref{fig3:Figure3Label}A2), and a high variability of the inter-spike-interval series (Figure \ref{fig3:Figure3Label}A3). It can also be deduced from raster plots of network activity for different levels of AR and different input frequency, shown in Figure \ref{fig3:Figure3Label}D.\\
\indent~The observation that signals of different periodicity are differentially processed by networks with different levels of asynchronous release suggests that an imposed level of AR might constrain a local circuit to optimally detect certain types of stimuli. To understand why such an optimization should occur, let us consider the simplest case of a single synaptic terminal that is stimulated by pulses at the rate $R$. To obtain an exact expression for the steady state rate of AR, we would need to solve the transcendental equation for the synaptic calcium variable: $\gamma R log(\frac{c_{0}}{c})=\frac{\beta c^{2}}{K_{c}^{2}+c^{2}}-I_{p}$. It is easy to see that the solution $c(R)$ of this equation is a monotonically increasing function of stimulation rate, $R$ and at least over some range can be thought of as being linear. The rate of AR, $\eta(c)$, is also a monotonically increasing function of synaptic residual calcium. Hence, the composite function $\eta(c(R))$ is itself a monotonically increasing function of $R$, and at least qualitatively can be approximated by the form  $\eta_{ss}(R)\propto \eta_{max}\frac{R^{4}}{R_{0}^{4}+R^{4}}$. The steady state value of the recovered resource variable $X$ (obtained under the conditions $\dot{X},\dot{Y},\dot{Z}=0$) is $X_{ss}=(1+(UR+\xi\eta(R))(\tau_{D}+\tau_{R}))^{-1}$. Then, the synaptic drive due to the asynchronous release at the model synapse is
\begin{equation}\label{eq:Eq_AsynchDrive}
Y_{a}(R)\propto X_{ss}(R)\eta_{ss}(R)=(1+(UR+\xi\eta(R))(\tau_{D}+\tau_{R}))^{-1}\eta_{max}\frac{R^{4}}{R_{0}^{4}+R^{4}}
\end{equation}
In Figure \ref{fig3:Figure3Label}B1, the steady state value of recovered resource (dashed line) is plotted along the two examples of low and high AR rates. As Figure \ref{fig3:Figure3Label}B2 shows, $Y_{a}(R)$ for the two examples shown in Fig. \ref{fig3:Figure3Label}B1 peaks at a certain frequency that is controlled by the parameters of synaptic depression. Simulation of a model synapse stimulated by a constant rate also confirms that AR peaks at a certain input rate (Fig. \ref{fig3:Figure3Label}B3). Note that if the noise is decoupled from the activity (does not depend on $R$), no such optimization with respect to the input rate is possible.\\
\indent~Going from the single synapse to the network level, a plot of coherence vs. $\nu_{in}$ (Fig. \ref{fig3:Figure3Label}C1) confirms that, for a given level of AR, the COS measure peaks at a certain input frequency, $\nu^{*}_{in}$. Yet, contrary to what is obtained for a \textit{single synapse} in Eq.\ref{eq:Eq_AsynchDrive} and in the related numerical calculations (Fig. \ref{fig3:Figure3Label}B3), higher $\eta_{max}$ appear to optimize the \textit{network} for higher $\nu_{in}$. Thus, AR  works together with feedback interaction between neurons to impart the network with stimulus selectivity. The firing rate (Figure \ref{fig3:Figure3Label}C2) and its variability (Figure \ref{fig3:Figure3Label}C3) are minimal at the optimal input rate (as compared with the responses for other input rates).\\
\indent~To sum up, in the periodic signal scenario, a neuronal network with plastic synaptic connections and activity-dependent asynchronous release of neurotransmitter appears to display selectivity with respect to the specific stimulus frequency. The preferred signal frequency, that produces maximally coherent output, depends on the maximal rate of asynchronous transmission at model synapses. In a sense, then, by modulating the rate of AR it is possible to modulate the "attention" of a network to different (periodic) stimuli.

\subsection{Response to frequency-modulated signals}
\begin{figure}[btp!]
\centerline{
\resizebox{0.7\textwidth}{!}{\includegraphics{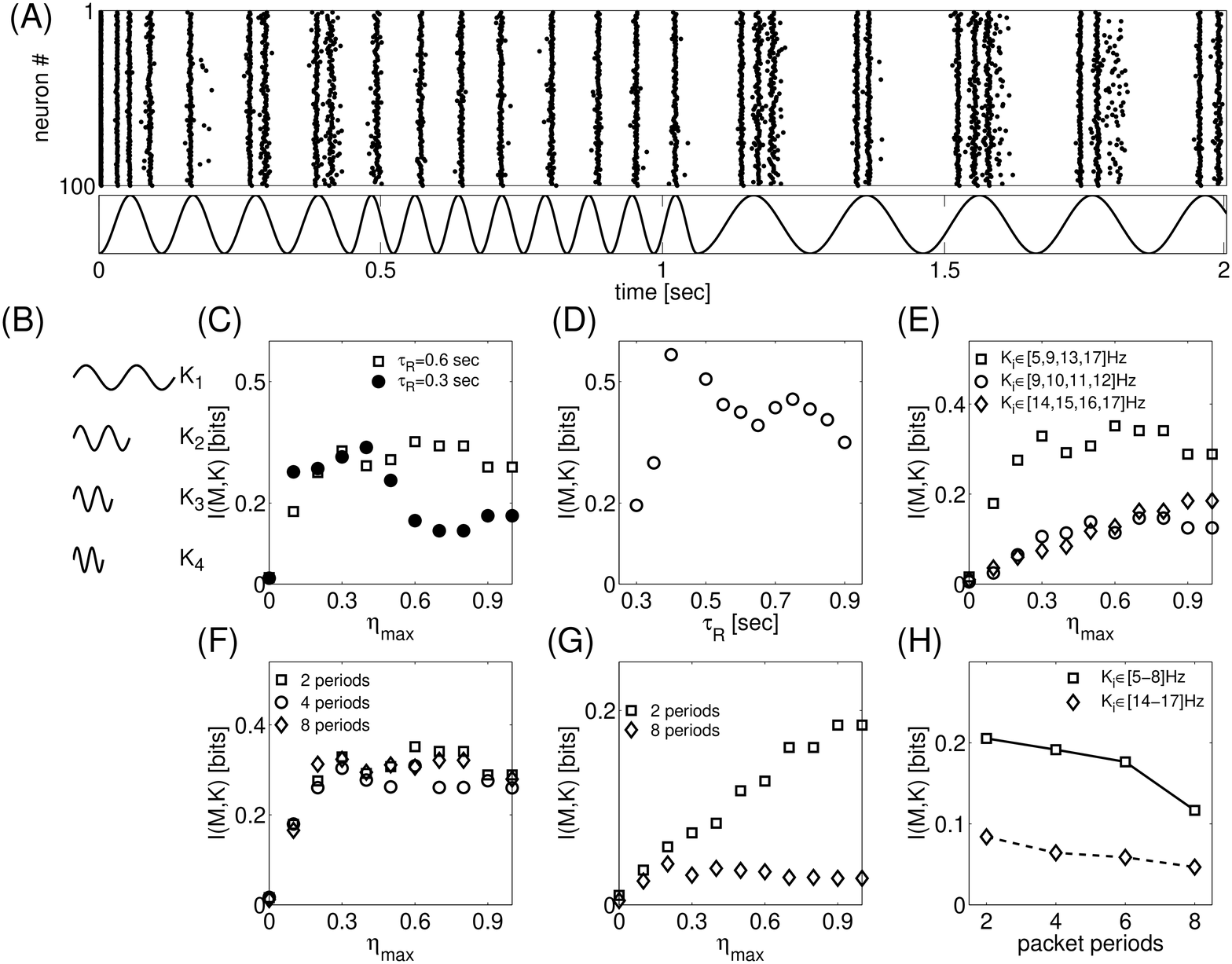}}}
\caption{\footnotesize Response to frequency-modulated signals. A) Raster plot of the network's activity in response to the frequency-modulated weak stimulus (lower panel). B) Examples of elementary packets of different frequencies from which the signal was constructed. The packet was defined as 2 periods of a sinusoidal wave. Here, $K_{1}=5~Hz, K_{2}=9~Hz, K_{3}=13~Hz, K_{4}=17~Hz$. C) Mutual information, $I(M,K)$, between input frequency packets and averaged ISI during that packet. Open squares: $\tau_{R}=0.6~sec$. Closed circles: $\tau_{R}=0.3~sec$. D) Mutual information plotted vs. recovery time from depression, for the case $\eta_{max}=0.6$. E) Mutual information plotted vs. the maximal rate of AR, for signal of different band-width (shown in legends). F) Mutual information for the case $K_{i}\in[5,9,13,17]Hz$, for different structural scenarios constructed as described in the text. G) Mutual information for the case $K_{i}\in[14,15,16,17]Hz$, for different structural scenarios constructed as described in the text. H) Mutual information vs. the number of packet periods (from less structured to more structured), plotted for two signals of different composition, for $\eta_{max}=0.4$. For all cases (E-H), $\tau_{R}=0.6~sec$.}
\label{fig4:Figure4Label}
\end{figure}

\indent~We have argued that a local recurrent circuit with activity-dependent synaptic connections and asynchronous release can exhibit selectivity with respect to the frequency of a periodic signal. However, real signals very rarely are purely periodic; rather, information that is carried by the signal is often encoded as amplitude modulation (AM), frequency modulation (FM), or both (AFM) \cite{DePitta08}. The case of amplitude modulation is quite problematic for a stochastic-resonance-like system, because by definition such a system is sensitive to the amplitude of the signal, so \textit{a priori} different components will not be treated equally. On the other hand, adding frequency modulation allows for the investigation of complex signal detection while keeping the amplitude fixed. To construct a frequency-modulated signal in a controlled way, we define signal packets, $K_{i}$, as 2-periods of constant amplitude but different frequency, $\nu_{i}$ (Figure \ref{fig4:Figure4Label}B). The frequency is fixed during the packet. The frequency-modulated signal is then represented as a series of packets, $\{K\}$ (Fig.\ref{fig4:Figure4Label}A).\\
\indent~Figure \ref{fig4:Figure4Label}A shows that the response of a network to such a composite signal depends on the packet's identity (frequency) as well as on the temporal context of its appearance. To what extent is it possible to predict the response of a network given stimulation by a certain packet? To quantify the ability of an input (packet) to reliably evoke a stereotypic response, we computed the mutual information between the input packets, $K_{i}$, and the pattern of activity that they evoke in a typical model neuron. Since neurons in our model typically fire several action potentials in response to a packet (Figure \ref{fig4:Figure4Label}A), we computed, for every appearance of each packet, the averaged inter-spike interval, $M=\{\langle ISI \rangle\}$, of neuronal activity during that packet. In so doing, we obtain two vectors that characterize the input (vector of packets, $K$) and the output (vector of packets-induced averaged ISIs, $M$). Mutual information is then computed as-
\begin{equation}\label{eq:Eq_MutualInfo}
I(M,K)=\sum_{\{K\}}\sum_{\{M\}}p(M,K)log_{2}(\frac{p(M,K)}{p(M)p(K)})
\end{equation}
where $p(K)$ is the probability to generate a $K$-th packet, $p(M)$ is the probability to observe averaged ISI that is equal to $M$, and $p(M,K)$ is the probability to observe averaged ISI that is equal to $M$ following stimulation by $K$-th packet. Since $\langle ISI\rangle$ (over a packet) fluctuates in time, in practice the $M$ values are distributed into 20 bins of equal size (each bin is $10-15$ msec) covering the range from smallest to largest observed ISIs.\\
\indent~Figure \ref{fig4:Figure4Label}C shows the dependence of the mutual information measure for the system driven by a composite signal, when $K_{i}\in[5,9,13,17] Hz$. For $\eta_{max}=0$ (no AR) $I(M,K)$ attains very low values, suggesting that the background noise $I_{bg}(t)$ in itself is not sufficient to reliably detect weak stimuli. Yet, increasing levels of asynchronous release enhance the detection of stimulus-related information. For high values of AR, this constructive effect of the noise is enhanced by stronger synaptic depression (larger $\tau_{R}$) (Figures \ref{fig4:Figure4Label}C,D). In particular, as is the case with the periodic signals, there seems to exist an optimal value of the synaptic recovery time at which $I(M,K)$ is maximized. We can intuitively understand why such an optimization should occur: in mildly-depressed networks, synapses recovery relatively quickly, and therefore the effect of AR is strong, leading to low predictability of output based on input. For large $\tau_{R}$, the destructive effect of depression becomes pronounced \cite{Fuhrmann02}.\\
\indent~Next, we wanted to assess the effects of signal structure on its detectability by our recurrent network. First, we compared input-output mutual information for signals with different band structure. As Figure \ref{fig4:Figure4Label}E shows, wide-band composite signals yielded higher values of $I(M,K)$ for all levels of AR than did narrow-band signals. Because different packets in the wide-band scenario are strongly separated (in terms of their frequency), the responses of a network to them are also strongly different. Therefore, it is less likely to make an erroneous identification of a packet. This is a nonlinear analog of the standard result for linear detection systems that the information rate is proportional to the signal bandwidth. In an additional series of tests, we probed the response of a network to correlated stimuli. Correlation was introduced here as a number of adjacent periods for packet's presentation to a network. As is seen from Figure \ref{fig4:Figure4Label}, correlation acted to reduce the mutual information, with the strongest effect exhibited for high-frequency narrow-band signal.\\

\section{Discussion}
Earlier theoretical studies of synaptic information transfer properties suggested that at high input rates the information carried by the synapse should be reduced because of the action of short-term presynaptic depression \cite{Fuhrmann02}. In this view, depression introduces strong fluctuations in the post-synaptic response, and therefore the mutual information between input and output (or the ability to reliably predict an output for a given input) is reduced. In our recurrent network, reducing the effect of depression (by making the recovery from it faster) resulted in the decrease of mutual information for high levels of asynchronous release. This suggests that while asynchronous release can prove beneficial for signal detection, in itself it is not sufficient, and relatively strong depression is needed in order to reveal the potential contribution of AR to the mutual information between input and output. Thus, in the noise-based scenario of signal processing, depression can actually play a constructive, rather than destructive, role.\\
\indent~Frequency selectivity was shown to emerge in feed-forward networks due to the opposing action of synaptic depression (that acts as a low-pass filter) and facilitation (that acts as a high-pass filter) \cite{Izhikevich03,Drover07}. Our findings are consistent with these earlier reports, but differ in several points. First, the facilitated AR in our model contributes to weak signal detection, thus providing us with the self-consistent framework of signal detection and processing. Second, feedback interactions due to network connectivity and slow time-scale of AR (that would introduce memory about earlier activation of different pathways) are expected to play an important role here. Earlier, we have shown that the size of the local circuit and its connectivity rule can play an important role in signal detection \cite{Volman08}. It would be interesting to see how AR shapes the function-form relation between the type of the signal (function), and the optimal network structure (form) to detect that signal.\\
\indent~Our observations that the coherence of spiking exhibits optimality with respect to the stimulus frequency may have broad consequences for signal processing by local neuronal circuits, provided that there exists a biophysical mechanism that would allow regulation of AR levels at synaptic terminals. One candidate is astrocyte-synapse cross-talk. Astrocytes, a sub-type of glial cells, are known to release chemical substances (glio-transmitters) that are able to exert regulatory effects on the pre-synaptic side by binding to different membrane receptors and activating Ca-associated down-stream processes \cite{Araque99}. Notably, a single astrocyte can "touch" $O(10^{4})$ synaptic terminals, thus potentially being able to coordinate the release properties of adjacent synapses \cite{Bushong02}. Theoretical studies (see, e.g. \cite{Volman07a,Nadkarni08}) have suggested that such astrocyte-synapse cross-talk might have important implications for brain function. In particular, in the recent study it was shown that feedback from astrocytes can "tune" the synapse to an optimal release probability in response to a periodic series of spikes \cite{Nadkarni08}. This optimal point of synaptic operation depended on the strength of the astrocyte-induced Calcium signal to the synapse. In the present work, the optimal signal frequency (for which the spiking coherence is maximal) depends on the level of AR, but the effect of AR is ultimately mediated by pre-synaptic Calcium. Thus, insofar as the explicit \textit{network-level} modeling of astrocyte-synapse interaction can be performed, our conclusions could be looked at as an extension of more detailed \textit{single synapse} studies reported in \cite{Nadkarni08}.\\
\indent~For higher-frequency narrow-band composite signals, prolonged presentations of different packets led to a dramatic fall in the ability of a system to reliably predict the pattern of input. This is consistent with the picture in which the action of synaptic depression (following long high-frequency stimulation) is to decrease the mutual information between input and output \cite{Fuhrmann02}. In a broader perspective, it suggests that different internal signal characteristics, such as its bandwidth and temporal structure (informational content) can affect its processing by the local neural circuit. In this view, information is no longer passively carried by the signal; rather, it actively determines the efficiency with which it will be detected by the target. The exact principles by which the structure of the information-carrying signals in such a system is constrained to yield its optimum are still elusive.\\

\section*{Acknowledgements: }We thank Eshel Ben-Jacob for insightful conversations. This research has been supported by the NSF-sponsored Center for Theoretical Biological Physics  (grant no. NSF PHY-0822283.)\\


\end{document}